\documentstyle[prl,aps,epsfig,preprint]{revtex}

\begin{document}
\draft
%\flushbottom

\title{Galactic Anisotropy as Signature of ``Top-Down'' Mechanisms of
Ultra-High Energy Cosmic Rays}

\author{S. L. Dubovsky and
P. G. Tinyakov} \address{Institute for Nuclear Research of the Russian
Academy of Sciences, 60th October Anniversary prospect, 7a, Moscow
117312, Russia.}  \maketitle

\begin{abstract}
{We show that ``top-down'' mechanisms of Ultra-High Energy Cosmic Rays
which involve heavy relic particle-like objects predict Galactic
anisotropy of highest energy cosmic rays at the level of minimum $\sim
20\%$. This anisotropy is large enough to be either observed or ruled
out in the next generation of experiments. }
\end{abstract}

%\pacs{98.70.Sa, 95.35.td}
%\narrowtext

The origin and nature of Ultra-High Energy Cosmic Rays (UHE CR) with
energies above $10^{19}$ eV \cite{UHE} is one of the actively debated
issues in modern astrophysics. At such a high energy the mean free
path of protons, nuclei and photons is much shorter than the size of
the Universe due to their interaction with cosmic microwave and radio
background \cite{GZK,photons}. Protons with $E > E_{\scriptsize\rm
GZK}\sim 5\times 10^{19}$ eV can only come from distances less than
$R_{\scriptsize\rm GZK}\sim 50$ Mpc; the corresponding region for
nuclei and photons is even smaller. The energy spectrum of cosmic rays
is thus expected to have a rapid falloff at energies $E\sim
E_{\scriptsize\rm GZK}$ (the so-called GZK cutoff).

The observation of few events with energies exceeding $10^{20}$ eV
seems to indicate that noticeable fraction of UHE CR comes from
relatively nearby ($R<R_{\scriptsize\rm GZK}$) sources. The latter is
not easy to reconcile with conventional astrophysical mechanisms
\cite{acceleration} as at these energies and distances the cosmic rays
are not substantially deflected by magnetic fields and point in the
direction of their source \cite{deflection}. Corresponding
astrophysical sources have not yet been identified. This prompts to
consider particle physics mechanisms of UHE CR production, which are
usually referred to as ``top-down'' mechanisms. Several such
mechanisms involving topological defects \cite{topolog} and decays of
primordial heavy particles \cite{ellis,X-KR,X-Ber,X-KT} have been
proposed. In this Letter we show that large class of the ``top-down''
mechanisms predict significant excess of highest energy events in the
direction towards the center of our Galaxy. Thus, these mechanisms can
be tested experimentally by studying the asymmetry in the angular
distribution of UHE CR. The predicted asymmetry is large enough to be
either observed or excluded by future experiments \cite{experiments}.

Our consideration concerns mechanisms in which production of UHE CR
involves heavy relic particle-like objects which behave as part of
cold dark matter (CDM). We call such mechanisms CDM-related. The
crucial feature of these mechanisms is that the distribution of
sources of UHE CR in the Universe between galaxies and intergalactic
space is proportional to that of CDM (cf. ref.\cite{cluster}) and does
not depend on their nature. In particular, the sources are mainly
concentrated in galactic halos, so that their average densities in the
Universe ($\bar n$) and in galactic halo ($\bar n_h$) are related by
\begin{equation}
{\bar n \over \bar n_h} \simeq {\Omega_{\scriptsize\rm CDM}\,
\rho_{\scriptsize\rm crit} \over
\bar\rho_{\scriptsize\rm halo} } \sim 10^{-5}.
\label{n/n_h}
\end{equation}
On the contrary, the distribution $n(x)$ of the sources in the
galactic halo depends on their interaction with each other and with
other matter and does not necessarily follow that of CDM\footnote{We
thank V.A.Kuzmin for drawing our attention to this point.}. For the
decay-type mechanisms $n(x)$ is simply the density of decaying
particles. For mechanisms based on collisions (see
e.g. \cite{monopoles}) $n(x)\propto \tilde n^2(x)$, where $\tilde
n(x)$ is the density of colliding particles. Note that in
the latter case the distribution $n(x)$ is typically more concentrated
around the galactic center.

The observed flux of UHE CR can be divided into Galactic and
extragalactic parts, 
\[
j = j_{\scriptsize\rm ext} + j_{h},
\]
where
\begin{equation}
j_h = C \int_{\scriptsize\rm halo} {d^3x \over x^2} n(x)
\label{j-halo}
\end{equation}
is the contribution of our Galaxy and 
\[
j_{\scriptsize\rm ext} = C \, 4\pi R_{\scriptsize\rm ext} \bar n
\]
has extragalactic origin. Here $R_{ext}=R_{\scriptsize\rm Universe}
\sim 4$~Gpc for energies below $E_{\scriptsize\rm GZK}$
and $R_{\scriptsize\rm ext}\sim 50$~Mpc for energies above
$E_{\scriptsize\rm GZK}$. The constant $C$ is the same in both
equations. Following ref.\cite{X-Ber} we note that eq.(\ref{n/n_h})
allows to estimate the relative magnitude of these two contributions,
\begin{equation}
{j_{\scriptsize\rm ext} \over j_h} =  
\alpha {R_{\scriptsize\rm ext} \over R_h} { \bar n \over \bar
n_h} \sim \alpha {R_{\scriptsize\rm ext} \over R_h} \times 10^{-5} ,  
\label{jext/jh}
\end{equation}
where $R_h\sim 100$~kpc is the size of the Galactic halo and $\alpha$
is the constant of purely geometrical origin,
\begin{equation}
\alpha = { 3\int_{r<R_h} d^3x n(x)\over R_h^2
\int_{r<R_h} {d^3x \over x^2} n(x)} . 
\label{alpha}
\end{equation}
Here $r(x,\theta)=(x^2+r_0^2-2xr_0\cos\theta)^{1/2}$ is the distance
between current point and the Galactic center while $r_0=8.5$ kpc
is the distance to the Galactic center.

Numerical value of $\alpha$ is close to 1 for uniform distribution
$n(x)=\mbox{const}$ and can be small for distributions concentrated
around the galactic center. Although $n(x)$ does not have to coincide
with CDM distribution in the halo, we consider as two examples 
the isothermal halo model \cite{halo1}
\begin{equation}
n(r)  \propto  {1\over (r_c^2 +r^2)}
\label{n=1/r^2}
\end{equation}
and more realistic distribution of ref.\cite{halo2}, 
\begin{equation}
n(r) \propto  {1\over \sqrt{(r_c^2 +r^2)} (R_h+r)^2},
\label{n=1/r}
\end{equation}
which we have arbitrarily regularized at $r=0$ by introducing the core
size $r_c$. The value of $\alpha$ is $\alpha\simeq 0.15$ and
$\alpha\simeq 0.5$ for distributions (\ref{n=1/r^2}) and
(\ref{n=1/r}), respectively, with no strong dependence on $r_c$ in the
range $r_c = 2 - 10$ kpc.

{}From eq.(\ref{jext/jh}) we find
\begin{eqnarray}
{j_{\scriptsize\rm ext} \over j_h} &\sim& \alpha
\makebox[1in]{}\mbox{for $E<E_{\scriptsize\rm GZK}$}, \nonumber \\ 
{j_{\scriptsize\rm ext} \over j_h}
&\sim& 10^{-2} \times \alpha
\makebox[1.2cm]{}\mbox{for $E>E_{\scriptsize\rm GZK}$}.  
\label{jext/jh-estimate}
\end{eqnarray}
Therefore, at $E<E_{\scriptsize\rm GZK}$ the Galactic and
extragalactic contributions can be comparable (although the Galactic
one is probably somewhat larger), while at $E>E_{\scriptsize\rm GZK}$
the extragalactic part is suppressed by a factor $\sim 10^{-2}$. In
either case a substantial fraction of the observed UHE CR should come
from the halo of our Galaxy. In this respect our conclusions agree
with that of ref.\cite{X-Ber}.

The Galactic part of the total UHE CR flux, $j_h$, is anisotropic due
to our position at 8.5 kpc from the center of the Galaxy. The
anisotropy can be obtained from eq.(\ref{j-halo}),
\[
j_h(\theta) \propto \int dx \, n(r(x,\theta)).
\]
Fig.1a shows the anisotropy $j_h(0)/j_h(\pi)$ as a function of the
core radius for the trial distributions (\ref{n=1/r^2}) and
(\ref{n=1/r}). For $n(x)=\mbox{const}$ the anisotropy is minimum and
constitutes about 20\%.  Fig.1b shows corresponding angular
dependencies of $j_h(\theta)$ at $r_c=5$ kpc. As can be seen from the
picture, the anisotropy of the galactic contribution is at least $\sim
20$\% and can be much larger if $n(x)$ is concentrated around the
galactic center. Also, it should be noted that the anisotropy depends
exclusively on $n(x)$ and does not depend on energy since cosmic rays
with energy $E\sim E_{\scriptsize\rm GZK}$ are deflected by the
Galactic magnetic field by only $\sim 3^{\circ}$ \cite{deflection}.

In turn, the extragalactic contribution $j_{\scriptsize\rm ext}$
consists of the isotropic part coming from distances $R\gg 50$~Mpc,
which is comparable in magnitude to $j_h$ and is present only at
$E<E_{\scriptsize\rm GZK}$, and the contribution from our
``neighborhood'' $R\lesssim 50$~Mpc. The latter should have peaks in
the direction of nearby galaxies and clusters. The contribution of
such a peak, $\delta j_{\scriptsize\rm ext}$, equals
\[
{\delta j_{\scriptsize\rm ext}\over j_h} =  \alpha
{R_h^2\over 3 R^2} {M\over M_G}, 
\]
where $R$ is the distance to the astronomical object, $M$ is its mass,
and $M_G$ is the mass of our Galaxy including halo. For instance,
contributions from Andromeda Nebula and Virgo Cluster are comparable
and close to $10^{-2}\times \alpha$, in agreement with
eq.(\ref{jext/jh-estimate}) and ref.\cite{X-Ber}.

Since at energies above the GZK cutoff the extragalactic contribution
is negligible, non-observation of the anisotropy at the level of $\sim
20$\% would rule out the CDM-related mechanisms of UHE CR. The
observation of the Galactic anisotropy would allow to reconstruct the
density profile $n(x)$ and, possibly, the distribution of CDM in the
Galactic halo. 

At energies below the GZK cutoff, the anisotropy is smaller due to the
relative enhancement of the isotropic extragalactic part. Since
anisotropy does not depend on energy and can be measured at
$E>E_{\scriptsize\rm GZK}$, it is possible, in principle, to determine
the magnitude of the extragalactic contribution. Provided the
CDM-related mechanisms are dominant at $E\lesssim E_{\scriptsize\rm
GZK}$ and the coefficient $\alpha$ is known, the ratio
$j_h/j_{\scriptsize\rm ext}$ could give, in view of eqs.(\ref{n/n_h})
and (\ref{jext/jh}), an important information about the distribution
of matter in the Universe.

Current data is not enough to draw definite conclusions about the
angular distribution of highest energy cosmic rays both because
of very limited statistics and the absence of data in the South
hemisphere where the Galactic center is situated. However, since the
anisotropy predicted by the CDM-related mechanisms is large, it will
be either observed or excluded already in the next generation of
experiments \cite{experiments}. Among these the Pierre Auger Project
has the best potential due to large number of expected events
(600--1000 events with $E>10^{20}$ eV in 10 years) and the ability to
see both hemispheres.

\acknowledgments{The authors are grateful to D.S.~Gor\-bu\-nov,
V.~A.~Kuz\-min, V.~A.~Rubakov, M.V.~Sazhin and D.V.~Semikoz for
helpful discussions. The work is supported in part by Award
No. RP1-187 of the U.S. Civilian Research \& Development Foundation
for the Independent States of the Former Soviet Union (CRDF), and by
Russian Foundation for Basic Research, grants 96-02-17804a and
96-02-17449a.}

\newpage

\begin{figure}[ht]
\begin{center}
\epsfig{file=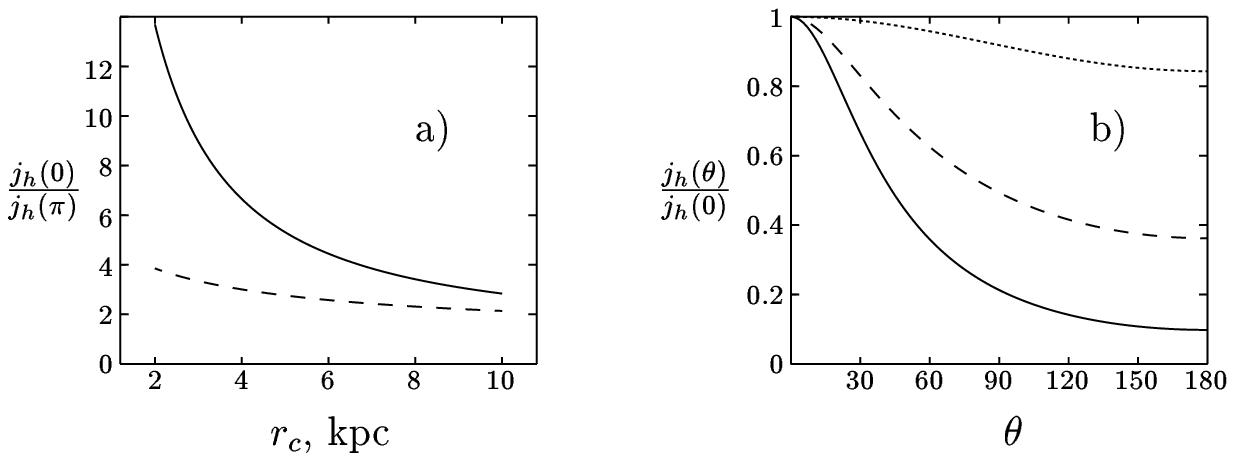,%
bbllx=120pt,bblly=475pt,%
bburx=490pt,bbury=635pt,
%width=236pt,height=175pt,%
clip=}
\end{center}
\caption{a)~The anisotropy $j_h(0)/j_h(\pi)$ as a function of the core
size $r_c$ for the density profiles (\protect\ref{n=1/r^2}) (solid
line) and (\protect\ref{n=1/r}) (dashed line). b)~The corresponding
angular distributions at $r_c=5$~kpc. The dotted line shows the
angular distribution for $n(x)=\mbox{const}$ (i.e., when the
anisotropy is minimum).}
\end{figure}

\iffalse
\begin{figure}[ht]
\begin{center}
\epsfig{file=fig1.ps,%
bbllx=100pt,bblly=535pt,%
bburx=257pt,bbury=695pt,
%width=236pt,height=175pt,%
clip=}
\end{center}
\begin{center}
\epsfig{file=fig1.ps,%
bbllx=285pt,bblly=535pt,%
bburx=460pt,bbury=695pt,
%width=236pt,height=175pt,%
clip=}
\end{center}
\caption{a)~The anisotropy $j_h(0)/j_h(\pi)$ as a function of the core
size $r_c$ for the density profiles (\protect\ref{n=1/r^2}) (solid
line) and (\protect\ref{n=1/r}) (dashed line). b)~The corresponding
angular distributions at $r_c=5$~kpc. The dotted line shows the
angular distribution for $n(x)=\mbox{const}$ (i.e., when the
anisotropy is minimum).}
\end{figure}
\fi

\end{document}